\documentclass{article}
\usepackage{graphicx,psfig,epsfig,rotating}
\def\title{\begin{center}\Large\bf}
\def\author(s){\vspace{0.3cm}\large\rm}
\def\text{\end{center}}

\newcommand{\degree}{\ensuremath{^{\circ}}}
\newcommand{\kms}{{\rm km~s}^{-1}}
\newcommand{\ha}{H$\alpha$}
\begin{document}

\title
Detections of extended galactic haloes of Virgo Cluster Galaxies\\

\author(s)
A.\ C.\ Katsiyannis$^{\rm 1}$, P.\ Boumis$^{\rm 2,3}$, J.\ Meaburn$^{\rm 4}$\\
\vspace{0.2cm}
$^{\rm 1}${\it Department of Pure and Applied Physics, 
           Queen's University Belfast, Belfast, BT7 1NN, UK}\\
$^{\rm 2}${\it University of Crete, P.O Box 2208, GR-71003, Heraklion, 
           Greece}\\
$^{\rm 3}${\it Foundation for Research and Technology-Hellas, P.O. Box 1527,
GR-71110 Heraklion, Greece}\\
$^{\rm 4}${\it Jodrell Bank Observatory, Department of Physics \& Astronomy, 
           University of Manchester, Macclesfield, Cheshire SK11 9DL}\\
\text

\vspace{0.3cm}

\large

\section*{Abstract}
The UK Schmidt camera was used to observe the Virgo cluster of
galaxies, where a total of 13 $R$-band Kodak Tech-Pan films were
obtained. The latter, were scanned by the APM machine (Cambridge, UK),
digitally aligned, co-added, corrected for vignetting effects and
finally cleaned of stellar features. The resulting image covers an
area of $\approx 6.2 \times 6.2$ degree$^{2}$ with a resolution of
$\approx 2$ arcsec pixel$^{-1}$, where extended objects with a surface
brightness up to 28 mag arcsec$^{-2}$ can been detected. Some of those
results have already be published in previous papers. Here we wish to
present a choice of faint detections and discuss some interesting
implications of this project.

\section{Introduction}

The actual dimensions of the galactic haloes is still an open
debate. Many authors including Kemp \& Meaburn (1991a, b, 1993, 1994,
1995), Kemp (1994) and reference therein, have used sky-limited
$B$-band photographic plates obtained from the United Kingdom 1.2-m
Schmidt Telescope (UKST) of the Anglo-Australian observatory (AAO), to
successfully detect extensive haloes of galaxies. More recently
Katsiyannis et al.\ (1998, 2000, 2001) used newly developed digitized
techniques to detect faint haloes of Virgo cluster galaxies utilizing
the whole 6.2\degree $\times$ 6.2\degree\ field of the UKST. To
increase the Signal-to-Noise ratio (S/N), 13 $R$-band low noise Kodak
Technical Pan (TecPan) films were co-added achieving detections of
haloes with surface brightness down to 28 mag arcsec$^{2}$. They
detected a number of extensive galactic haloes as well as filaments,
details of galactic interactions, etc. This paper presents detections
of more extensive halos of well known Virgo cluster galaxies.

\section{Extensive haloes}

M 60 (NGC 4649) is a giant elliptical galaxy with an S0(2) E2
morphological type (Arp 1968; Binggeli et al. 1985; Caon \& Einasto
1995) and a heliocentric velocity of $\sim 1095~\kms$~(Knapp et al.\
1989; Faber et al.\ 1989). It is located in the eastern extension of
the Virgo cluster, about 1 Mpc from the cluster centre (Trinchieri et
al.\ 1997a). Figure 1 contains the M 60 galaxy in two different
contrasts. The left-hand-side image is M 60 under normal contrast
while in the right-hand-side is the same galaxy under the highest
possible contrast. The dimensions of the halo are $\approx 17 \times
14~{\rm arcmin}^{2}$~and assuming a distance of 17 Mpc (Mould et al.\
1995) for all the Virgo cluster members we can derive the physical
dimensions of this halo as $\approx 85 \times 70~{\rm kpc}^{2}$. The
apparant proximity with the NGC 4647 galaxy can be easily seen but
there is no evidence in our data of any interactions between the two
objects. De Vaucouleur et al.\ (1976) reported total magnitude of
$B_{T}= 9.83$ while the total X-ray luminosity (ROSAT surveys) is
$\sim 1.5$~to $1.8 \times 10^{41}$ erg s$^{-1}$~(Trinchieri et al.\
1997b; Irwin \& Sarazin 1998; Beuing et al.\ 1999). The radio and
\ha~luminosities are $L_{R} \sim 10^{38}$ erg s$^{-1}$ (Stanger \&
Warwick 1986) and $L_{H\alpha}= 6.79 \times 10^{39}$ erg s$^{-1}$
(Trinchieri et al.\ 1991), respectively.

\begin{figure}
\centering
\mbox{\epsfclipon\epsfxsize=12.5cm\epsfbox[48 301 547 541]{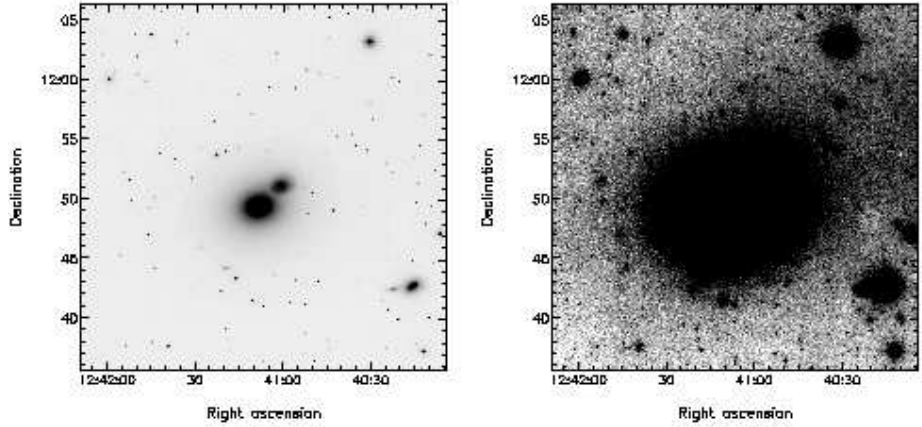}}
\caption{Deep extended image of M 60 in normal (left) and high (right) 
         contrast. A halo of $\approx 85 \times 70~{\rm kpc}^{2}$ can
         be clearly seen in the latter.}
\label{fig1}
\end{figure}

NGC 4429, an early-type lenticular galaxy (S0/Sapec, Sandage \&
Tammann 1981; de Vaucouleur et al.\ 1991) is located in a group of
$\sim 60$ galaxies with NGC 4472 at the group centre (Garcia
1993). Its total magnitude is $B_{T} = 11.15$~ (Bingelli et al.\ 1985)
and the X-ray luminosity $9.1 \times 10^{39}$ erg s$^{-1}$~(Burstein
et al.\ 1997), while the radio flux density was measured by Niklas et
al.\ (1995) to $S_{6.3 {\rm cm}} \approx 1$~mJy.  In Figure 2,
NGC4429's extensive halo is clearly visible. The left-hand-side image
(Fig.\ 2a) is the galaxy under normal contrast, while the
right-hand-side image is the same target at the highest possible
contrast. The detected halo is $\approx 8.3 \times 3.5~{\rm
arcmin}^{2}$ or $\approx 41 \times 17~{\rm kpc}^{2}$.

\begin{figure}
\centering
\mbox{\epsfclipon\epsfxsize=14.5cm\epsfbox[48 340 529 541]{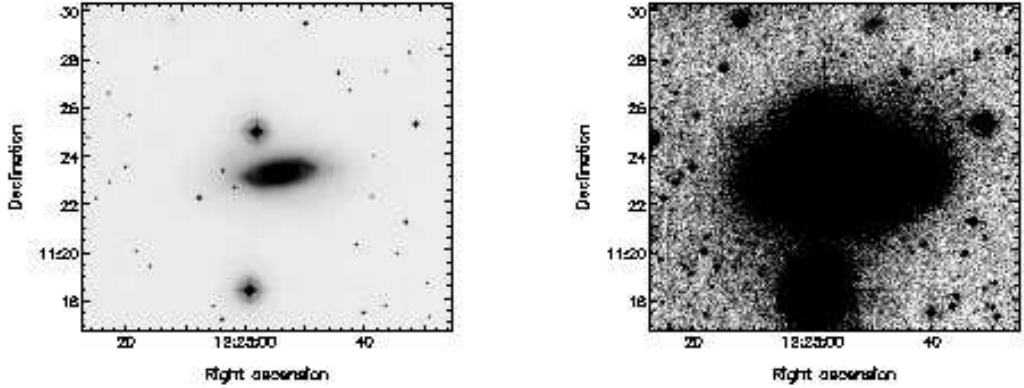}}
\caption{Deep extended image of NGC 4429. The images shown on the left 
         and right hand-side are of normal and high contrast, 
         respectively. The dimensions of the halo are $\approx 41 
         \times 17~{\rm kpc}^{2}$.}
\label{fig2}
\end{figure}

NGC 4473 is a highly elongated elliptical galaxy. Morphologically it
is classified as E5 by Morton \& Chevalier (1973) while Jaffe et al.\
(1994) classified NGC 4473 as Type II (`disky' elliptical galaxy)
based on the galaxy's shape and brightness profile. Recently,
Macchetto et al.\ (1996) and Ferrari et al.\ (1999) have shown a very
small edge-on dust disk. Figure 3 presents another extensive halo
detection. The normal contrast image (Fig.\ 3a) of NGC 4473 is
$\approx 2 \times 4~{\rm arcmin}^{2}$ ($\approx 10 \times 20~{\rm
kpc}^{2}$), while the high contrast image reveals dimensions of
$\approx 4.1 \times 6.3~{\rm arcmin}^{2}$ ($\approx 20 \times 30~{\rm
kpc}^{2}$). Garcia (1993) has argued that NGC 4473 is located in a
group of 25 galaxies within the `Virgo II' cloud (named `Southern
Cloud II') and with NGC 4501 to be at the centre of this group. One
unusual feature of this galaxy is it's high redshift of $2236 \pm
197~\kms$~(Young et al.\ 1978) as the average velocity of the cluster
and its velocity dispersion is $110 \pm 20~\kms$ (Williams 1977). Head
et al.\ (1976) suggested a total magnitude of $11.08 \pm 0.05$ while
the absolute $B$-band magnitudes for the $B$~and $V$~are $-20.5 \pm
0.5$~(Malumuth \& Kirshner 1985) and $-22.57$~(Gebhardt et al.\ 1996),
respectively.

\begin{figure}
\centering
\mbox{\epsfclipon\epsfxsize=14.5cm\epsfbox[48 321 534 541]{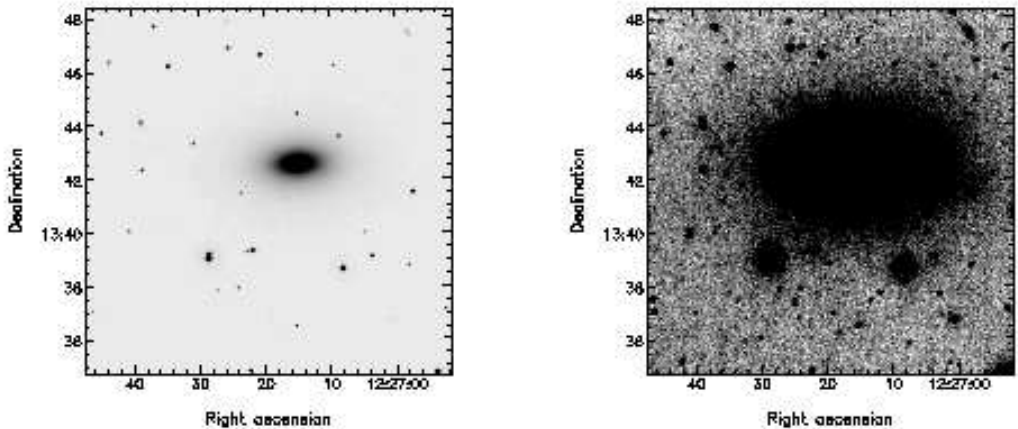}}
\caption{Normal and high contrast images of NGC 4473. The estimated 
         halo dimensions are $\approx 20 \times 30~{\rm kpc}^{2}$.}
\label{fig3}
\end{figure}

\section{Conclusion}

Previous publications have discussed many interesting conclusions
concerning this project. The authors would like to emphasize the
detection of previously unobserved extended galactic material in
almost all of the $\approx$20 galaxies studied so far. As these
detections were noise limited, we have no reason to believe that the
majority of these detected haloes are not even further
extended. Although our sample dealt mainly with the Virgo cluster, we
believe that this important conclusion applies to most of the galaxies
with similar angular dimensions.

\section*{Acknowledgments}
P.\ Boumis acknowledges support from a $``$P.EN.E.D." 
program of the General Secretariat of Research and Technology of
Greece.

\section*{References}

\noindent Arp H.; 1968; PASP; 80; 129

\noindent Beuing J., Dobereiner S., Bohringer H., Bender R.; 1999; MNRAS; 302; 209

\noindent Binggeli B., Sandage A., Tammann G.A.; 1985; AJ; 90; 1681

\noindent Burstein D., Jones C., Forman W., Marston A.P., Marzke R.O.; 1997; ApJS; 111; 163

\noindent Caon N., Einasto M.; 1995; MNRAS; 273; 913

\noindent de Vaucouleurs G., de Vaucouleurs A., Corwin H.G. Jr.; 1976; Second Reference Catalogue of Bright Galaxies, University of Texas, Austin

\noindent de Vaucouleurs G., de Vaucouleurs A., Corwin H.G. Jr., Buta R., Paturel G., Fouque P.; 1991; Third Reference Catalogue of Bright Galaxies, Springer-Verlag, New York

\noindent Faber S.M., Wegner G., Burstein D., Davies R.L., Dressler A., Lynden-Bell D., Terlevich R.J.; 1989; ApJS; 69; 763

\noindent Ferrari F., Pastoriza M.G., Macchetto F., Caon N.; 1999; A\&AS; 136; 269

\noindent Garcia A.M.; 1993; A\&AS; 100; 47

\noindent Irwin J.A., Sarazin C.L.; 1998; ApJ; 499; 650

\noindent Jaffe W., Ford H.C., O'Connell R.W., Van den Bosch F.C., Ferrarese L.; 1994; AJ; 108; 1567

\noindent Katsiyannis A.C., Boumis P., Meaburn J.; 2000; in IAU Symp. 205; Galaxies and their constituents at the highest angular resolutions, eds. R. T. Schilizzi et al. (ASP Press), 122

\noindent Katsiyannis A.C., Boumis P., Meaburn J.; 2002; A\&A; in preparation

\noindent Katsiyannis A.C., Kemp S.N., Berry D.S., Meaburn J.; 1998; A\&AS; 132; 387

\noindent Katsiyannis A.C., Kemp S.N., Berry D.S., Meaburn J.; 2001; Ap\&SS; 276; 733

\noindent Kemp S.N.; A\&A, 282, 425

\noindent Kemp S.N., Meaburn J.; 1991a; MNRAS; 251; 10

\noindent Kemp S.N., Meaburn J.; 1991b; MNRAS; 252; 27

\noindent Kemp S.N., Meaburn J.; 1993; A\&A, 274, 19

\noindent Kemp S.N., Meaburn J.; 1994; A\&A; 289; 39

\noindent Kemp S.N., Meaburn J.; 1995; in ASP Conference Series Volume 84, `The Future Utilisation of Schmidt Telescopes', eds., J. Chapman, R. Cannon, S. Harrison, B. Hidayat, p. 200 (San Francisco, Ca)

\noindent Knapp G.R., Guhathakurta P., Kim D-W., Jura M.; 1989; ApJS; 70; 329

\noindent Macchetto F., Pastoriza M., Caon N., Sparks W.B., Giavalisco M., Bender R., Capaccioli M.; 1996; A\&AS; 120; 463

\noindent Morton D.C., Chevalier R.A.; 1973; ApJ; 179; 55

\noindent Mould J., Aaronson M., Huchra J.; 1980; ApJ; 238; 458

\noindent Niklas S., Klein U., Wielebinski R.; 1995; A\&A; 293; 56

\noindent Sandage A., Tammann G.A.; 1981; A Revised Shapley-Ames Catalog of Bright Galaxies (Carnegie Institution of Washington, Washington, DC)

\noindent Stanger V.J., Warwick R.S.; 1986; MNRAS; 220; 363

\noindent Trinchieri G., di Serego Alighieri S.; 1991; AJ; 101; 1647

\noindent Trinchieri G., Fabbiano G., Kim D-W.; 1997; A\&A; 318; 361

\noindent Trinchieri G., Noris L., di Serego Alighieri S.; 1997b; A\&A; 326; 565

\noindent Williams T.B.; 1977; ApJ; 214; 685

\noindent Young P., Sargent W.L.W., Boksenberg A., Lynds C.R., Hartwick F.D.A.; 1978; ApJ; 222; 450

\end{document}